\documentstyle [12pt]{article}
\newcommand{\ben}{\begin{enumerate}}
\newcommand{\een}{\end{enumerate}}
\newcommand{\be}{\begin{equation}}
\newcommand{\ee}{\end{equation}}
\newcommand{\bse}{\begin{subequation}}
\newcommand{\ese}{\end{subequation}}
\newcommand{\bea}{\begin{eqnarray}}
\newcommand{\eea}{\end{eqnarray}}
\newcommand{\bc}{\begin{center}}
\newcommand{\ec}{\end{center}}

\newcommand{\IR}{\mbox{I \hspace{-0.2cm}R}}

\author
{
{\bf  Mohammad Vahid TAKOOK \thanks{e-mail: takook@ccr.jussieu.fr}}\\
{\it  Departement of physics, Razi University, }\\
{\it  Kermanshah, IRAN}\\
}
\title{Covariant two-point function for \\
        linear gravity in de Sitter space}
\begin{document}
\maketitle
\begin{abstract}
The Wightman two-point function for the gravitational field in the linear 
approximation 
(the rank-2 ``massless'' tensor field) on de Sitter space has a 
pathological behaviour for large
separated points (infrared divergence). This behaviour can be eliminated 
in the two-point function
for the traceless part of this field if one chooses the Gupta-Bleuler 
vacuum. But it is not
possible to do the same for the pure trace part (conformal sector).  We 
briefly
discuss the consequences of this pure trace behaviour for inflationary 
models.
\end{abstract}

\section{Introduction}
The graviton propagator on de Sitter (dS) space (in its usual linear 
approximation 
for the gravitational fields)  for large separated points has a 
pathological behaviour (infrared
divergence) [Allen, Turyn, $1987$; Floratos, Iliopoulos, Tomaras, $1986$; 
Antoniadis, Mottola,
$1991$]. Some authors proposed that infrared divergence could rather be 
exploited in order to create
instability of the dS universe [Ford, $1985$; Antoniadis, Iliopoulos, 
Tomaras, $1986$]. The field
operator for linear gravity  in dS space has been considered in this way 
by Tsamis and Woodard in
terms of flat coordinates which cover only one-half of the dS hyperboloid 
[Tsamis, Woodard, $1992$].
They have examined the possibility of quantum instability and they have 
found a quantum field which
breaks dS invariance. However, we show that this behaviour problem for 
the traceless part of the
field disappears if one uses the Gupta-Bleuler vacuum defined by [de 
Bievre, Renaud, $1998$; Gazeau,
Renaud, Takook,
$1999$]. On the other hand,  such a procedure is unsuccessful for the 
pure-trace part of the field
(conformal sector). In the general relativity framework, one cannot 
associate a dynamics to the
conformal sector because the physical content of this field is not 
apparent. It is coordinate or
gauge dependent. Therefore one may think that its behaviour troublesome 
may originate from imposing
the gauge invariance and has no actual physical consequence. But, in the 
presence of a matter quantum
field, that part of the metric acquires a dynamical content and the 
problem appears in any attempt
to quantize it.

In a previous paper, we have shown that  one can write the rank-2 
``massive'' tensor field 
(divergencelesse or ``transverse'' and traceless) in terms of a 
projection operator and a scalar
field. At the ``massless'' limit,  there appears a singularity in the
projection operator. This type of singularity appears precisely because 
of the divergenceless
condition. By dropping the divergenceless condition, we can make the 
mentioned singularity in the
tensor field (for its traceless part only) disappear. In quantizing this 
field, there appears another
singularity in the Wightman two point function like in the case of the 
``massless'' minimally
coupled scalar fields [Allen, Folacci,
$1987$]. The latter type of singularity appears because of the zero mode 
problem for the
Laplace-Beltrami operator on dS space. In order to solve it, we must 
follow the procedure
already used for a completely covariant quantization of the minimally 
coupled scalar field [Gazeau,
Renaud, Takook, 1999]. 

The organization of this paper is the following. Section $2$ is 
devoted to the traceless field and it is explained how the choice of the 
Gupta-Bleuler vacuum
eliminates pathological behaviour. In Section $3$ we examine the 
questions raised by the pure-trace
part. Section
$4$ is a brief conclusion on the inflationary universe scenario.

\section{Traceless part}

Here, we briefly recall our de Sitterian notations. The de Sitter 
space-time is made 
identical to the four dimensional one-sheeted  hyperboloid
\be X_H=\{x \in \IR^5 ;x^2=\eta_{\alpha\beta} x^\alpha  x^\beta 
=-H^{-2}\},\;\; \alpha,
\beta=0,1,2,3,4, \ee
where $\eta_{\alpha\beta}=$diag$(1,-1,-1,-1,-1)$. The de Sitter metrics is
\be  ds^2=\eta_{\alpha\beta}dx^{\alpha}dx^{\beta}=
g_{\mu\nu}^{dS}dX^{\mu}dX^{\nu},\;\; \mu=0,1,2,3,\ee
where $X^\mu$ are the $4$ space-time coordinates in dS hyperboloid.
We use the tensor field notation $K_{\alpha\beta}(x)$ with respect to the 
ambiant space, and the
transversality condition 
$ x.K(x)=0$ is imposed.  In this notation, it is simpler to express the 
tensor field (and also the
two-point function) in terms of scalar fields.

The two-point function for the ``massive'' spin-$2$ field 
$K_{\alpha\beta}^{tt}(x)$ 
(``transverse'' or divergenceless and traceless) is defined by [Gazeau, 
Takook]
$$ {\cal W}_{\alpha\beta \alpha'\beta'}(x,x')=\langle 
\Omega,K_{\alpha\beta}^{tt}(x)K_{\alpha'\beta'}^{tt}(x')\Omega  \rangle $$
\be {\cal W}_{\alpha\beta \alpha'\beta'}(x,x')=D_{\alpha\beta 
\alpha'\beta'}^{tt}(x,x')
{\cal W}(x,x'). \ee
${\cal W}(x,x')$  is the Wightman  two-point function  for the massive 
scalar field on dS space. \newline
$D_{\alpha\beta \alpha'\beta'}^{tt}(x,x')$ is a 
projection tensor, which
satisfies the ``divergencelesse'' and traceless conditions. In the limit of 
the  ``massless'' spin-$2$
field there appear two types of singularity in the two-point function. 
The first one lies in the
projection tensor $D_{\alpha\beta\alpha'\beta'}^{tt}(x,x')$ and it 
disappears if one fixes the gauge
(the dropping of the divergenceless condition).  The other one lies in 
the scalar Wightman two-point
function ${\cal W}(x,x')$ (the minimally coupled scalar field) and it 
disappears if we follow the
procedure presented in [Gazeau, Renaud, Takook, $1999$]. Then the 
two-point function is defined by
[Gazeau, Renaud, Takook] 
$$ {\cal W}_{\alpha\beta \alpha'\beta'}(x,x')=\langle 
\Omega,K_{\alpha\beta}^{t}(x)
K_{\alpha'\beta'}^{t}(x')\Omega  \rangle $$
\be {\cal W}_{\alpha\beta \alpha'\beta'}(x,x')=
\Delta_{\alpha\beta\alpha'\beta'}^t
(x,\partial;x',\partial'){\cal W}(x,x'), \ee
where $ \Delta^t(x,\partial;x',\partial')$ is a projection tensor which 
satisfies the 
traceless condition. ${\cal W}$ is the two-point function for the 
minimally coupled scalar field in
the Gupta-Bleuler vacuum [Takook, $1997$]  
\be {\cal W}(x,x')=\frac{iH^2}{4\pi} \epsilon (x^0-x'^0)[\delta(1-{\cal 
Z}(x,x'))-\theta 
({\cal Z}(x,x')-1)], \ee
where $ {\cal Z}=-H^2 x.x'$ and
$ \epsilon (x^0-x'^0)=\left\{\begin{array}{rcl} 1&x^0>x'^0\\ 0&x^0=x'^0\\ 
-1&x^0<x'^0.\\ 
\end{array}\right.$ 

\section{Conformal sector}

The tensor field that we considered in the previous section is traceless. 
But in the general 
case the
tensor field consists of a traceless part and a pure-trace part 
(conformal sector):
\be K_{\alpha\beta}(x)=K_{\alpha\beta}^{t}(x)+K_{\alpha\beta}^{pt}(x).\ee
The pure trace part can be written in the form
$$K_{\alpha\beta}^{pt}(x)=\frac{1}{4}\theta_{\alpha\beta} \psi, $$
where $\psi$ is scalar field and $\theta_{\alpha \beta}=\eta_{\alpha 
\beta}+ 
H^2x_{\alpha}x_{\beta}$. With a certain choice in the gauge condition , 
we are able to write down
the following field equation for the scalar field $\psi$ [Gazeau, Renaud, 
Takook]
\be (\Box_H-5H^2)\psi=0.\ee
So this field cannot be interpreted in terms of a unitary irreducible 
representation of the dS group.
Difficulties arise when we want to quantize such fields which show 
negative squared mass in their
wave equation. The corresponding two-point functions have a pathological
large-distance behaviour (infrared divergence)[Gazeau, Renaud, Takook]. 
We just emphasize on the
fact that, so far, this degree of freedom should not appear as a physical 
one. 

\section{Conclusion}

We conclude that the pathological large-distance behaviour for the 
physical degree 
of freedom of the linear gravity in the Wightman two-point function can 
be easily cured. Antoniadis,
Iliopoulos and Tomaras have also  shown that the pathological 
large-distance behaviour of the
graviton propagator on a dS background does not manifest itself in the 
quadratic part of the
effective action in the one-loop approximation [Antoniadis, Iliopoulos 
and Tomaras; $1996$]. That
means that this behaviour may be gauge dependent and it should not appear 
in an effective way in a
physical quantity. On the other hand, it exists in an irreducible way in 
the pure-trace part
(conformal sector). The conformal sector may be interesting for 
inflationary universe scenarii. In
these theories, one introduces an inflaton scalar field. Because of this 
field, the conformal sector
of the metric becomes dynamical and it must be quantized [Antoniadis, 
Mazure, Mottola, $1997$]. Then
it produces a gravitational instability. This gravitational instability 
and the primordial quantum
fluctuation of the inflaton scalar field define the inflationary model. 
The latter can explain the
formation of the galaxies, clusters of galaxies and the large scale 
structure of the universe
[Lesgourgues, Polarski, Starobinsky,
$1998$].

We may conclude that the quantum instability of dS space and the breaking 
of the dS invariance are both due to the quantization of the conformal 
sector.

{\bf  Acknowlegements}
\noindent We are grateful to J-P. Gazeau J. Iliopoulos and J. Renaud for 
very useful discussions.



\begin{thebibliography}{a}
\addcontentsline{toc}{chapter}{Bibliographie}

\bibitem  -Allen B., Folacci A., {\it Phys. Rev.,} {\bf D} 35 (1987) 
3771  
\bibitem  -Allen B., Turyn M., {\it Nucl. Phys.,} {\bf B} 292 (1987) 813
\bibitem  -Antoniadis I., Iliopoulos  J., Tomaras T. N., {\it Phys. Rev. 
Letters }  56 (1986) 1319 
\bibitem  -Antoniadis I., Iliopoulos  J., Tomaras T. N., {\it Nucl. 
Phys.,} {\bf B} 462 (1996) 437 
\bibitem  -Antoniadis I., Mottola  E., {\it Phys. Rev.,} {\bf D} 45 
(1991) 2013 
\bibitem -Antoniadis I., Mazur P.O., Mottola  E., {\it Phys. Rev.,} {\bf 
D} 55 (1997) 4770
\bibitem -De Bi\`evre S.,  Renaud J., {\it Phys. Rev.,} {\bf D} 57 (1998) 
6230
\bibitem  -Floratos  E. G., Iliopoulos J., Tomaras T. N., {\it Phys. 
Letters,} {\bf B} 197 (1987) 373 
\bibitem  -Ford  H. L., {\it Phys. Rev.,} {\bf D} 31 (1985) 710  
\bibitem  -Gazeau J. P.,  Renaud J., Takook M.V., gr-qc/9904023 appear in 
class. quantum Grav.
\bibitem  -Gazeau J. P., Takook M.V.,  in preparation {\it ``Massive'' 
spin-$2$ field in the de Sitter universe}
\bibitem  -Gazeau J. P.,  Renaud J., Takook M.V., in preparation {\it 
linear covariant quantum gravity in de Sitter space} 
\bibitem  -Lesgourgues, Polarski, Starobinsky, astro-ph/9807019 
\bibitem  -Takook M.V., Th\`ese de l'universit\'e Paris VI,   1997 {\it 
Th\'eorie quantique des
champs pour des syst\`emes \'el\'ementaires ``massifs'' et de ``masse 
nulle'' sur l'espace- temps de
de Sitter.}
\bibitem  -Tsamis N. C., Woodard R. P., {\it Phys. Letters,} {\bf B} 292 
(1992) 269 

\end{thebibliography}
\end{document}